# Vehicle as a Resource (VaaR)

Sherin Abdel Hamid[1]      Hossam S. Hassanein[1]      Glen Takahara[2]

[1]School of Computing, Queen's University, Kingston, ON, Canada

[2]Department of Mathematics and Statistics, Queen's University, Kingston, ON, Canada

{sherin, hossam}@cs.queensu.ca         takahara@mast.queensu.ca

## ABSTRACT

Intelligent vehicles are considered key enablers for intelligent transportation systems. They are equipped with resources/components to enable services for vehicle occupants, other vehicles on the road, and third party recipients. In-vehicle sensors, communication modules, and on-board units with computing and storage capabilities allow the intelligent vehicle to work as a mobile service provider of sensing, data storage, computing, cloud, data relaying, infotainment, and localization services. In this paper, we introduce the concept of Vehicle as a Resource (VaaR) and shed light on the services a vehicle can potentially provide on the road or parked. We anticipate that an intelligent vehicle can be a significant service provider in a variety of situations, including emergency scenarios.

## I.     INTRODUCTION

With high demand for reducing the number of vehicular fatalities and enhancing ITS applications and services, many newly-manufactured vehicles will be equipped with components that will classify them as 'intelligent vehicles'. Such components include sensors and actuators with intra-vehicle communication, and electronic control units (ECUs) for processing and operation control. Vehicles will be equipped with a wireless communication module for supporting three types of communication: 1) between vehicles, known as vehicle-to-vehicle (V2V) communication, 2) between vehicles and infrastructure (V2I and I2V), or 3) between vehicles and any neighboring object (V2X). In addition, an On-Board Unit (OBU) will be integrated in each vehicle for interaction with drivers, displaying warnings, issuing alerts, offering automotive services/infotainment, and managing the communication with a vehicle's surroundings. Powerful OBUs can be considered in-vehicle PCs that can handle computing tasks supported by abundant storage capabilities. Although currently partially available in some luxury models, the availability of such components will be expanded to most vehicles in the near future. Pivotal components are shown in Figure 1.

With these components, a vehicle can be considered a mobile resource for many services such as sensing, data relaying and storage, computing, cloud, infotainment, and localization. We introduce the concept of *Vehicle as a Resource (VaaR)*, which focuses on such vehicular potential. The VaaR vision is stimulated by the ubiquity of vehicles (with predictable mobility patterns) and the vehicular resources that are readily available. The abundance of such on-board resources distinguishes the use of a vehicle as a resource from other mobile resource providers, viz. smartphones, which suffer from limited resources and lack of trajectory prediction. We anticipate that a vehicle will be a mobile provider for diversified resources currently unimagined and will be a key enabler for the revolution of the Internet of Things (IoT). Vehicles on a road or at a parking lot with idle resources and capabilities can cooperatively form a powerful resource for services that can benefit a wide scope of service domains.

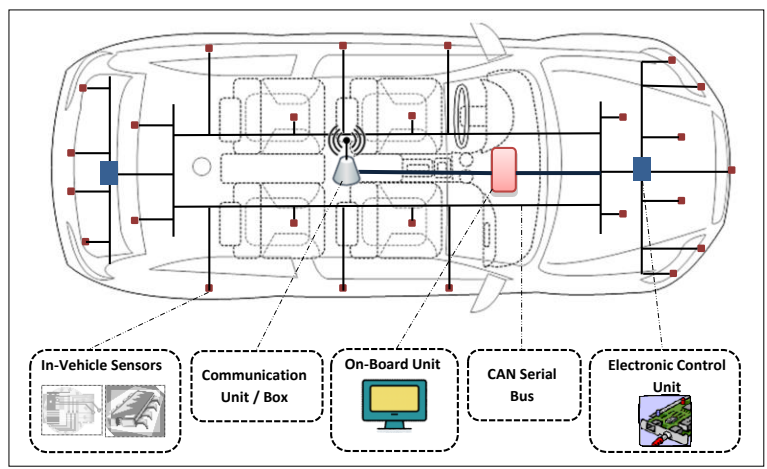

**Figure 1 In-vehicle components shaping the intelligent vehicle.**

Another enabler of VaaR is the availability of a multiplicity of wireless communication technologies for communication between a vehicle and its surroundings. One such technology specifically introduced for vehicular use is the Wireless Access for Vehicular Environment (WAVE) technology which is based on the IEEE 802.11p standard and Dedicated Short Range Communication (DSRC). Another is the Communication Access for Land Mobiles (CALM) standard. CALM will support an integrated communication unit that provides many air interfaces that include 2G/3G cellular, infrared, millimeter-wave, mobile wireless broadband (HC-SDMA, 802.16e [WIMAX/WiBro] and 802.20), satellite, and DSRC. In addition, some Zigbee communication modules are designed to support vehicular communication. As well, Visible Light Communication (VLC) is now gaining high interest, and many systems have been proposed for the vehicular environment. Having such technologies available in a vehicle will provide flexible communication with its surroundings regardless of the type of air interface available.

In this paper, we delineate the diversified resources a vehicle can provide as instances of VaaR. As well, we elaborate on how these resources can be tapped into and pooled for performing certain tasks or providing certain services. In addition, for each resource, we highlight some potential applications/services that become feasible with the aid of using a vehicle as a provider for this resource. We focus on the services provided by the vehicle to other vehicles and third parties, as these have the greatest potential for widely expanded benefits. We will also discuss some prominent challenges that face the wide adoption of VaaR.

The remainder of the paper is organized as follows. In Section II, we introduce the VaaR concept and present the different resources a vehicle can provide in discussing the various instances of VaaR. Illustrative scenarios that show the benefit and applicability of VaaR are discussed in Section III. In Section IV, we discuss some challenges that may face the widespread use of VaaR and can be considered open points for research. Section V concludes the paper.

# II. VEHICLE AS A RESOURCE (VAAR)

## A. VaaR-Sensing

According to the automotive sensors market growth in North America, the average number of sensors per vehicle has reached 70 in 2013 [1] – 100 in some luxury vehicles. A vehicle can then be considered a significant resource of sensory data. We categorize sensors according to their application domain: 1) Safety, 2) Diagnostics, 3) Convenience, and 4) Environment Monitoring, as shown in Figure 2. *Safety sensors* are the most crucial being targeted at decreasing accidents and driving fatalities. *Diagnostics sensors* provide on-board detection of component malfunction to avoid further breakdowns or damage. In addition to on-board services, the system can include a reporting capability to support remote diagnostics. *Convenience sensors* support comfort and convenience applications for drivers and passengers. Most of these sensors are deployed inside the vehicle compartment to provide direct services for its occupants, while others are deployed for providing driving assistance. Finally, *environment monitoring sensors* provide alerts/warnings about road hazards or reports about traffic, road and weather conditions.

Having a variety of sensors along with communication capabilities shapes the concept of VaaR-Sensing. Many applications and platforms have been proposed to make use of vehicles as sensing resources, mainly for environment monitoring as their measured data can be of benefit beyond a vehicle's compartment. In these applications, vehicles sense/monitor the surrounding environment and store the sensed data for further relaying- either without processing or after processing to search for data of interest. Such data can be reported to third parties through the Internet or V2X (Vehicle-to-Any) communications. These third parties can be data centers that can offer the data for public/commercial services, or can be other mobile users/drivers. Such sensing services can expedite the adoption of public sensing in its participatory and opportunistic forms.

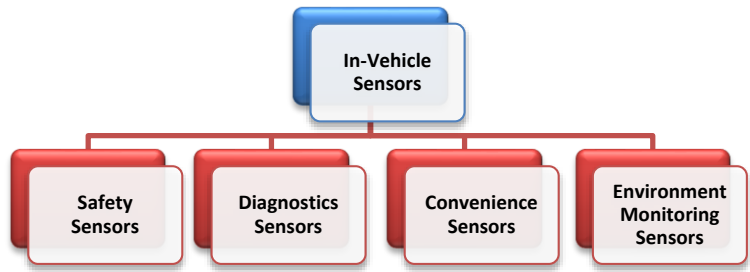

**Figure 2 Categories of in-vehicle sensors.**

An example of a vehicular sensing platform is the MobEyes platform [2] which utilizes vehicles as mobile sensors to monitor surroundings, recognize objects, store data, and advertise this data for potential sharing.

The MobEyes platform uses V2V communication. Other platforms are Internet-based, where vehicles collect and process sensor data locally then send the processed data to database servers through the Internet for further analysis and publishing. Examples of such platforms are discussed in [3].

Currently, there is high focus on utilizing in-vehicle sensors for providing road condition monitoring services. Many techniques have been proposed from using accelerometers to the use of cameras for capturing photos and videos that are further analyzed to extract road features. A prominent road monitoring platform is CarMote [4].

## *B. VaaR-Data Storage and Computing*

Data-storage and computing are tightly coupled as both services are offered through the on-board computers in intelligent vehicles. Advanced in-vehicle computers are currently available, some almost as powerful as personal computers, such as those in [5] with dual core processors up to 2.8GHz and storage capabilities in Gigabytes. Due to advances in data storage technologies, it is anticipated that in-vehicle storage capacity will reach multiples of Terabytes in the future, enabling the vehicle to act as a mobile data server.

With such powerful computing resources, it is foreseen that computing tasks would be offloaded to vehicles. Vehicles on a road or a parking lot can be considered a distributed system that can potentially manage computing tasks more efficiently and cost-effectively than a centralized computing center.

Many platforms are proposed to utilize vehicles as a resource for data storage for further diffusion or retrieval of the data by other agents. Vehicle-generated data or data obtained from neighboring vehicles can be stored until the vehicle reaches a dedicated data collector or kept in the vehicle until retrieved as a reply to queries sent by data-seeking vehicles/agents. This communication paradigm is considered a case of delay/disruption-tolerant networks.

The MobEyes platform mentioned earlier is an example of both VaaR-Sensing and VaaR-Data storage. Each vehicle keeps the sensed data in local storage, and advertises its stored data via meta-data. While moving, other vehicles harvest the meta-data and send queries to obtain data of interest. Another example is the FleaNet platform [6] that utilizes vehicles as information traders. In FleaNet, vehicles are used to resolve queries generated by other vehicles. While moving, vehicles receive queries and data advertisements from other vehicles and/or information generated by roadside advertisement stations (Adstations), and store this data/information locally. When vehicles receive queries, they try to resolve them by consulting their own local storage to provide a possible match.

Currently there is an emerging vision of utilizing intelligent vehicles for cloud services. With sensing, computing, and storage resources, and abundant power supply and communication modules, a vehicle can work as a powerful cloud. As with conventional clouds, the owner of a vehicle can rent out vehicular resources on demand when these resources are not in use.

Olariu et al. [7] introduced the term 'autonomous vehicular clouds' (AVCs), arguing that in-vehicle resources may be underutilized by traditional vehicular applications as motivation to 'take vehicular networks to the clouds'. They remark that the benefit of on-board resources will be maximized if resources of multiple vehicles are combined.

VaaR-Cloud has been proposed as a mobile experimental laboratory in areas with limited facilities that hinder the use of a remote service. Another scenario utilizes idle parked vehicles in a parking lot of a company as a distributed computing asset. Tasks can be offloaded to vehicles during the workday in lieu of building or renting an outsource infrastructure. Vehicle owners can be compensated so both the company and employees can benefit. Similar arguments can be applied to vehicles parked at an airport. Resources are ample and travellers' vehicles may be left unutilized for days, resulting in computing resources that, if managed properly, could turn an airport parking lot into a data center. Travellers can share their travel plans with the airport, which can then schedule and manage the resources. Parked vehicles that are part of the vehicular cloud would be plugged into a standard power supply and can be provided with an Ethernet connection.

Vehicular clouds can have advantages over fixed clouds in applications such as the mobile experimental laboratory. Another advantage of vehicular clouds is their autonomous formation. Neighboring vehicles can autonomously form a cloud to provide instantaneous services (e.g., collecting traffic information at congested intersections for traffic light management). A detailed comparative study of vehicular and conventional clouds is presented in [8].

## C. VaaR-Data Relaying

Wireless message exchange between vehicles is known as inter-vehicle communication and a network of communicating vehicles is known as a Vehicular Ad-Hoc Network (VANET). A VANET is a sub-category of wireless multi-hop networks in which a source depends on intermediate nodes to relay messages to a destination. In this communication paradigm, vehicles can be considered a resource for relaying data to other nodes out of the communication range of the source node. A vehicle can be used not only for relaying data to other neighboring vehicles but also to/from Road Side Units (RSUs).

As an example of VaaR-Data Relaying, Delay-Bounded Vehicular Data Gathering (DB-VDG) [9] is a solution that supports geographical data gathering services and depends heavily on vehicles as data relays.

In DB-VDG, queries can be sent to retrieve information from areas of interest and replies can be routed back from vehicles in these areas. Both queries and replies are delivered via intermediate vehicles that act as relays. DB-VDG can also be considered an example of VaaR-Sensing as the gathered data is acquired from the in-vehicle sensors.

Vehicles can be good candidates for commercial advertisements that utilize vehicular mobility to disseminate the ads through relays. These commercial advertisements can be pushed to a vehicle either through the Internet or by Adstations. Examples include store special offers or restaurants menus.

VaaR-Data Relaying can also be a technique for extending network coverage as presented in [10]. The authors propose a vehicle-to-vehicle relay (V2VR) scheme to extend a road side access point (AP) service range and allow drive-thru vehicles to have an extended coverage range. In this scheme, when a vehicle approaches an AP, it selects a vehicle ahead to work as a relay of the AP traffic and allow early access to the AP services. When a vehicle is about to leave the direct AP coverage range, it selects another vehicle behind to work as a relay for the AP traffic and extend the access time while maintaining high throughput.

VaaR-Data Relaying can play a pivotal role in cases of emergency or natural disasters where infrastructure may be broken down. In these cases, vehicles can help deliver critical data to crisis/disaster-management authorities, collect environmental data, and inform emergency vehicles about optimal routes.

## D. VaaR-Infotainment

Infotainment refers to the combination of information with entertainment. With the on-board communication capabilities supporting communication with surrounding mobile agents (e.g., other vehicles, cellular phones, or communication-enabled handheld devices), and providing possibility for always-on Internet access, a vehicle can be a great source of infotainment.

Some vehicular infotainment services are solely Internet-dependent (e.g., web surfing, email access, video downloads and online gaming), and others are supported by inter-vehicle communications (e.g., exchanging information and files between vehicles). More detailed examples of these infotainment services can be found in [11]. The entertainment part of the infotainment resource is usually for the benefit of the vehicle's occupants and is not the focus of this paper, which is on the information part.

A vehicle can provide others with information obtained from its own sensors, other vehicles, RSUs, direct access to the Internet, or Adstations. This information may include traffic and environmental conditions, navigational and safety information, parking availability, or commercial advertisements. Many of the vehicular platforms mentioned earlier in the paper can be considered examples of VaaR-Infotainment when the utilization of the other vehicular resources is ultimately for the sake of providing information.

As most of the infotainment applications are real-time, such applications face challenges caused mainly by the highly dynamic vehicular topologies. These challenges must be dealt with to support the required QoS levels. The work in [12] presents some of these challenges along with proposed solutions and some issues that still need consideration.

## *E. VaaR- Localization*

Vehicles can be considered potential resources for locating objects. Through their sensing and communication capabilities, they can recognize and locate objects, and send this information. For instance, MobEyes can be used to enable vehicles to recognize the license plate numbers, store them, and broadcast representative meta-data. Other mobile agents (e.g., police patrol cars) can retrieve the recognized plate numbers to locate and track lost or stolen vehicles.

Vehicles can be used to locate neighboring vehicles for the sake of estimating the distance to these vehicles or informing them about their positions for accuracy purposes, etc. An example of this positioning ability is the Intel patent 'Visible Light Communications and Positioning (VLCP)' [13] for proposing the use of visible light communication as an inter-vehicle communication technique and introducing a scheme for positioning neighboring vehicles.

Vehicles can also use self-localization techniques to determine their own position, complementing and/or refining GPS information. These self-benefit services are out of the paper focus.

## III. VAAR IN ACTION

In this section, we discuss scenarios illustrating some of the various resources a vehicle can provide on the move. The main scenario depicted in Figure 3 shows an emergency scenario caused by an accident on the road.

Vehicles G and H had a head-on collision. Due to that collision, traffic has stopped moving in the vicinity of the collision. Vehicle D is a truck carrying goods that need to be delivered on time for shipping. The driver of the truck needs to know how severe the situation is. He sends a request using his vehicle's OBU asking for information about the collision from the neighbors of the involved vehicles. He receives a reply from vehicle A that it can provide live video streaming of the situation. He agrees and starts getting live video streaming from node A. Vehicles A and D are out of communication range of each other. Therefore, they depend on vehicles B and C to work as data relays for their shared content. In this scenario, vehicle A can be considered an example of VaaR-Infotainment, while vehicles B and C can be considered examples of VaaR-Data Relaying.

As a consequence of the collision, vehicle H had a malfunction in its GPS. The driver of vehicle H needs to know its location to call for help. He can depend on the resources of vehicle A to locate his vehicle. In this case, vehicle A can be considered an example of VaaR-Localization as well.

As a direct neighbor of vehicle G, vehicle F has recorded the collision using its on-board camera. It decides to store the recorded video for further reporting. Therefore, vehicle F can be considered an example of VaaR-Data Storage.

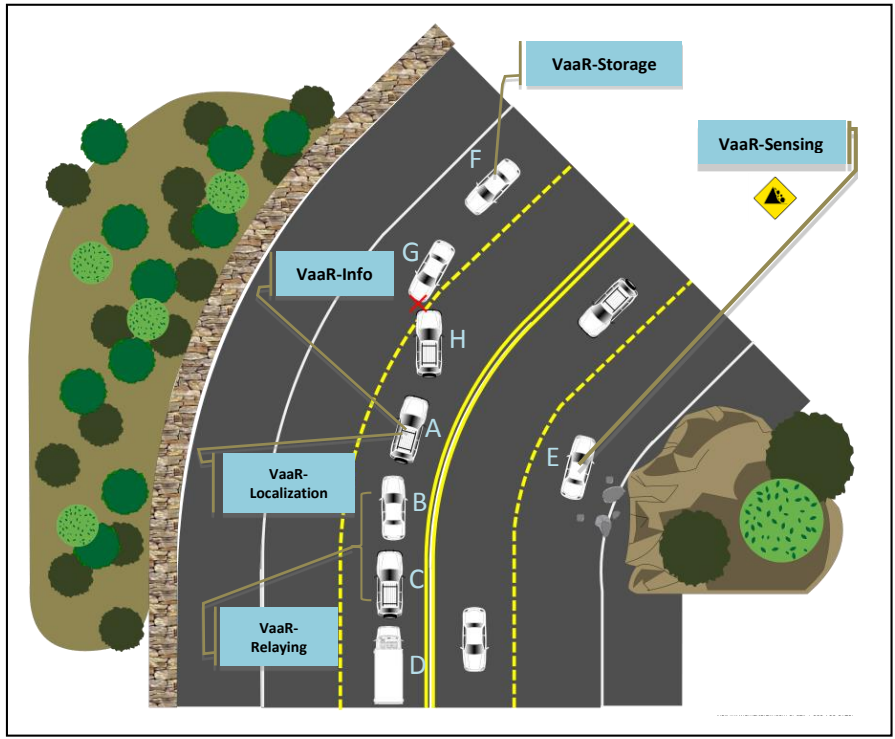

**Figure 3** Illustrative scenario showing the viability of VaaR. Vehicles *G* and *H* had an accident and vehicles *F*, *A*, *B*, and *C* work as resource providers while being in the vicinity of the emergency situation. Vehicle *E* as well works as a resource after detecting falling rocks on its way.

Finally, on the other side of the road, vehicle E has detected falling rocks using its in-vehicle sensors. It reports this hazard using its communication module to the municipality to take proper action to avoid accidents. This is considered an example of VaaR-Sensing.

## IV. CHALLENGES AND OPEN ISSUES

Although VaaR brings a wide scope of benefits to various applications, many challenges face its wide penetration and adoption. In this section, we highlight some of the prominent challenges and open issues. A summary of the challenges and some potential solutions can be found in Table I.

*-- Privacy, quality, and redundancy pose major challenges in collecting data from and offloading tasks to vehicular resources*

The open access and resources sharing that VaaR promises bring privacy to the forefront. Since a vehicle's resources may be accessible to many users to share, data privacy should be guaranteed for all users sharing these same resources. Privacy should be maintained for vehicles' owners as well. This issue can be handled by virtualization and scheduling techniques for coordinating access to shared resources. Other techniques can be used for protecting data and controlling its access to its owner only. Examples of such techniques include the use of personal data vaults, which are individually controlled data repositories. Privacy is needed as well to hide the identity of participants when they prefer to remain anonymous. Pseudonymity can be considered a solution that can help in hiding the actual identities for the sake of reducing the possibility of linking between the sent data and participants.

The quality of information (QoI) retrieved from vehicles is also a concern. Data/information retrieved needs to be verified and validated before making decisions and/or publishing to the public. As well, quality reports about cooperating vehicles and corresponding reputation scores should be maintained for future reference. Depending on the criticality of the supported services, such reputation scores can be computed based on either short- or long-term participation history with the aid of watchdog modules. Tasks can be offloaded to those participants with reputation scores higher than a certain threshold to guarantee a certain level of QoI. Feedback can be given to participants about the quality of their reported data and their perceived reputation levels.

While having a plethora of vehicles on the road increases the pool of resources, it may lead to having large amounts of redundant data reported resulting in waste of both communication and computation resources. Such redundancy can be eliminated by adopting data aggregation and fusion techniques. Vehicular clusters and chains can be formed on the road while electing a leader node to take care of the intra-cluster/chain data aggregation. As well, redundancy of the reported data can be avoided by applying proper selection and recruitment techniques that ensure selecting participants with minimal overlapping.

*-- The intermittent/dynamic availability of resources hinders extended usage*

Vehicles' mobility is considered as a plus that allows vehicles to cover wider areas compared to their static counterparts. However, while pooling resources from a vehicle, the vehicle may leave the area of interest or connectivity before reporting/relaying the desired data, or finishing the task in hand. Dynamic availability of resources may also be temporal stemming, for instance, from the high need for resources during the rush hours and their idleness during late nights. Effective resource management techniques are needed for proper task assignment and retrieval. A vehicle's cyber-physical existence should be taken into consideration in pooling/tapping into vehicular resources. Prediction techniques can help in anticipating the spatiotemporal availability span; hence, taking more informed allocation decisions. Task sharing and handover can also be potential solutions of the partial availability.

*-- Incentives are needed to encourage owners to offer their vehicular resources*

The VaaR concept cannot thrive without active user participation especially if users (vehicle owners) lose their willingness to participate and share resources. Some form of incentive and direct service/reward must be provided in return. Some of the incentive mechanisms proposed for P2P file sharing [14] can be applied for incentivizing vehicular users especially for those applications that have major commonalities with the P2P paradigm. In general, incentives can be of three types: 1) willingness to serve the public, 2) getting service in return, or 3) getting monetary returns. Earlier studies have shown that incentives with monetary values are the most effective. Monetary incentives can be in the form of pecuniary returns, vouchers, or passes. Another incentive for the owners of parked vehicles can be offering them free parking, while their idle resources are being accessed. The value of such incentives can be determined by the data collecting/task distributing party through the use of pricing models that can consider the level of participation and QoI, or through reverse auction techniques that are currently getting popular in the rewarding models.

*-- Recruitment mechanisms should be deployed for efficient selection of participants*

In urban environments, offloading tasks to vehicles to utilize their resources is challenging in terms of participant selection as there can be many potential participants in an area of interest that may reach a hundred of vehicles in a congested area. These participants cannot all be recruited for a task as this would result in a higher cost in terms of participants' rewards. Selecting the right number of potential participants in a cost-effective manner, while providing an acceptable level of service, is an open area. Such recruitment mechanisms must also consider the QoI challenge mentioned earlier. A scheme of that aims at minimizing the number of recruited participants while achieving a given level of coverage for an area of interest can be found in [15]. Although the recruitment schemes proposed for utilizing the resources of smartphones cannot be directly applied to vehicles due to the different design considerations, a study of potential adaptation of such schemes is worthwhile.

*-- How to power up the resources of parked vehicles for tasking is a technical challenge*

Parked vehicles considered for resource utilization should be plugged into a power supply. However, it is not logical to keep their PCs on all the time waiting for potential tasking. Solutions are needed to power these PCs only when required. These solutions can be categorized into 'on-demand' and 'pre-scheduled' techniques. An on-demand approach [16], requires ECUs and the in-vehicle PC are connected through the Controller Area Network (CAN) bus working as the main intra-vehicle communication bus. One of the features of CAN-connected nodes is that they can operate in a sleep mode such that while a vehicle is stopped, these nodes consume a minimum amount of battery. The CAN-connected in-vehicle PC of a parked vehicle can be remotely powered up on-demand. Experiments are needed though to test the feasibility of the wakeup/probing techniques. Other techniques that fit under the second category include deploying scheduling mechanisms to schedule the power up times a priori and sending this schedule whenever vehicles are on. Such techniques may also include assistance from mobility prediction mechanisms to anticipate a vehicle's daily parking times.

*-- Consolidation of different resources is expected and interoperability should be guaranteed*

It is expected that vehicles as resources of services will be integrated with other resource providers. For example, VaaR-Sensing may be merged with other sensing resources like the smartphones and sensor networks with each paradigm being assigned a part of a sensing task based on its availability and capabilities. Such consolidation of resources entails the development of communication, synchronization, and resource management techniques that ensure syntactic and semantic interoperability and seamless handover among the different paradigms. Sensor fusion techniques can be considered to handle the collaborative-sensing example mentioned above. Generally speaking, information fusion/integration techniques should be considered to handle merging information from different sources. Work proposed for the operation of the heterogeneous networks (HetNets) is promising [17].

## V. CONCLUSIONS

With its diversified on-board resources, an intelligent vehicle can be considered a mobile service provider that can assist in a wide scope of applications and domains. In this paper, we introduced the concept of Vehicle as a Resource (VaaR) to unveil the potential of an intelligent vehicle on the road or in a parking lot. We showed that a vehicle can be a resource for sensing, data storage, computing, cloud, data relaying, infotainment and a means for locating other objects. As well, we presented demonstrating scenarios, challenges, and open issues related to the adoption of VaaR. With the presented VaaR vision, we anticipate that an intelligent vehicle will get the information services and the intelligent transportation systems to an era of service revolution.

Table I  VaaR challenges and potential solutions

| Challenge | Potential Solutions |
|---|---|
| **Privacy** | <ul><li>Applying virtualization and scheduling techniques.</li><li>Use of personal data vaults.</li><li>Hiding identities through pseudonymity.</li></ul> |
| **Data quality** | <ul><li>Adopting verification and analysis techniques to build quality reports and history.</li><li>Computing reputation scores to be considered in selection refinement.</li><li>Giving feedback to participants about their data quality.</li></ul> |
| **Redundancy** | <ul><li>Can be eliminated by adopting data aggregation and fusion techniques while forming vehicular clusters and chains with representative nodes.</li><li>Can be avoided with proper participant selection.</li></ul> |
| **Dynamic availability of resources** | <ul><li>Adopting effective resource management techniques for proper task assignment and retrieval.</li><li>Adopting prediction techniques to anticipate the availability spatiotemporal span.</li><li>Considering task sharing and handover techniques.</li></ul> |
| **Incentives for offering resources** | <ul><li>Adopting P2P incentive mechanisms.</li><li>Offering vehicles' owners some rewards in return.</li><li>Focus on the encouraging monetary incentives and their corresponding pricing models.</li></ul> |
| **Participant selection** | <ul><li>Proposing efficient recruitment mechanisms that consider the rewards, budget, and data quality.</li><li>Studying the feasibility of adapting the smartphone recruitment schemes.</li></ul> |
| **Powering up parked vehicles PCs** | <ul><li>Utilizing features of the CAN bus connecting the intra-vehicular resources.</li><li>Use of scheduling and mobility prediction mechanisms.</li></ul> |
| **Consolidation of different resource providers** | <ul><li>Considering data fusion and information integration techniques to merge data/information from different sources.</li><li>Considering on-going research in the area of 'integration of heterogeneous networks.'</li></ul> |